\documentclass[aps,prl,preprint,superscriptaddress]{revtex4-1}

\usepackage{graphicx}
\usepackage{amsmath}
\usepackage{mathrsfs}
\usepackage{amssymb}
\usepackage{url}
\usepackage{color}
\usepackage{soul}
\usepackage{hyperref}
\usepackage{mathtools}
\usepackage[detect-none]{siunitx}

\usepackage{wasysym}

\begin{document}


\title{High-aspect-ratio silica meta-optics\\
for high-intensity structured light}

\author{Beatriz Oliveira}
\affiliation{Department of Physics, Instituto Superior Técnico, Universidade de Lisboa, Lisbon, Portugal}
\affiliation{Instituto de Engenharia de Sistemas e Computadores - Microsistemas e Nanotecnologias (INESC MN), Lisbon, Portugal}

\author{Pablo San Miguel Claveria}
\affiliation{GoLP/Instituto de Plasmas e Fusão Nuclear, Instituto Superior Técnico, Universidade de Lisboa, Lisbon, Portugal}

\author{Pedro D.R. Araujo}
\affiliation{Instituto de Engenharia de Sistemas e Computadores - Microsistemas e Nanotecnologias (INESC MN), Lisbon, Portugal}

\author{Patricia Estrela}
\affiliation{GoLP/Instituto de Plasmas e Fusão Nuclear, Instituto Superior Técnico, Universidade de Lisboa, Lisbon, Portugal}
\affiliation{Currently with the Laboratory for Laser Energetics, University of Rochester, Rochester, NY, USA}

\author{Ines Gonçalves}
\affiliation{Department of Physics, Instituto Superior Técnico, Universidade de Lisboa, Lisbon, Portugal}
\affiliation{Instituto de Engenharia de Sistemas e Computadores - Microsistemas e Nanotecnologias (INESC MN), Lisbon, Portugal}

\author{Maria Ines S. Nunes}
\affiliation{Department of Physics, Instituto Superior Técnico, Universidade de Lisboa, Lisbon, Portugal}
\affiliation{Instituto de Engenharia de Sistemas e Computadores - Microsistemas e Nanotecnologias (INESC MN), Lisbon, Portugal}

\author{Rui Meirinho}
\affiliation{Instituto de Engenharia de Sistemas e Computadores - Microsistemas e Nanotecnologias (INESC MN), Lisbon, Portugal}

\author{Marta Fajardo}
\affiliation{GoLP/Instituto de Plasmas e Fusão Nuclear, Instituto Superior Técnico, Universidade de Lisboa, Lisbon, Portugal}

\author{Marco Piccardo}
\email[]{marco.piccardo@tecnico.ulisboa.pt}
\affiliation{Department of Physics, Instituto Superior Técnico, Universidade de Lisboa, Lisbon, Portugal}
\affiliation{Instituto de Engenharia de Sistemas e Computadores - Microsistemas e Nanotecnologias (INESC MN), Lisbon, Portugal}

\maketitle 

\textbf{Structured light and high-intensity ultrafast lasers are two rapidly advancing frontiers in photonics, yet their intersection remains largely unexplored. While ultrafast lasers continue to push the boundaries of peak intensities, structured light has enabled unprecedented control over light’s spatial, temporal, and polarization properties. However, the lack of robust optical devices capable of bridging structured light with the high-intensity domain has constrained progress in combining these directions. Here, we demonstrate high-aspect-ratio silica meta-optics, which close this gap by combining silica’s extraordinary damage resistance with the advanced phase and polarization control offered by metasurfaces. By leveraging anisotropic etching techniques, we fabricate nanopillars exceeding 3 $\mu$m in height with aspect ratios up to 14, enabling precise manipulation of complex light fields at intensities far beyond the thresholds of conventional metasurfaces. We showcase their functionality in generating vortex beams and achieving polarization manipulation with large phase retardance at challenging long-visible wavelengths. High-aspect-ratio silica meta-optics unlock structured laser-matter interactions in extreme regimes, that will surpass plasma ionization thresholds and enable applications such as relativistic particle acceleration and high-harmonic generation with structured beams, for both tabletop ultrafast systems and large-scale laser facilities.}

\bigskip

\textbf{\large{Introduction}}

The rise of high-peak-power pulsed lasers\cite{Li2023,national2018opportunities} has revolutionized photonics, enabling breakthroughs from precision material processing with minimal thermal damage to ultrafast medical interventions within internal tissues and even particle acceleration\cite{Joshi1984} to relativistic energies. Despite these advances, most developments have focused on increasing pulse energy, while the laser beams themselves have remained largely conventional, dominated by standard linearly-polarized Gaussian modes. Meanwhile, a separate yet vibrant effort in structured light\cite{Forbes2021} has evolved over the past decades, bringing an unprecedented control over all its degrees of freedom—amplitude, phase, polarization, and time/frequency content—and establishing the paradigm of multimode photonics\cite{Piccardo2021,Wright2022,Wright2022b}, with spatiotemporal fields emerging as its most dynamic frontier\cite{Shen2023}. Although these capabilities have rarely been extended to high-intensity regimes, this gap highlights the opportunity to transfer the power of ultrashort, high-intensity lasers into the rich design space of structured light\cite{harrison2024progress}.

Metasurfaces offer a promising route to advanced beam shaping in compact, flat-optics platforms\cite{Yu2011,Yu2014,Arbabi2015}. By engineering arrays of subwavelength scatterers, one can impart virtually arbitrary phase and polarization transformations on an incident beam, making it possible to craft exotic forms of light\cite{Dorrah2022}. However, most metasurfaces employ different materials for the nanostructures and the substrate, creating interfaces that are particularly vulnerable in high-intensity scenarios. These heterogeneous interfaces, prone to nucleating laser-induced damage under large electric-field gradients\cite{Mangote2012}, have prevented conventional metasurfaces from being adopted in high-intensity laser applications.

Full-silica optics stand out as an ideal candidate for high-power applications due to their intrinsic robustness\cite{Mangote2012}, stemming from silica’s wide bandgap and the absence of material interfaces, which eliminate potential damage sites. Additionally, silica offers high optical transparency over a wide spectral window, encompassing common intense laser systems such as Ti:sapphire, Nd:glass, and their frequency-upconverted sources. However, the low refractive index of silica has so far restricted the design of these metasurfaces due to poor light confinement in the nanopillars, severely limiting their phase control. Current fabrication methods further constrain the degree of freedom over individual nanoelements. For example, large-area printing approaches based on laser dewetting of metallic masks\cite{Ray2021} can produce broadband anti-reflective features\cite{Ray2020} but lack control over the shape and orientation of individual scatterers. All-glass metalenses developed for applications in virtual reality and biological imaging have struggled to achieve the vertical sidewall profiles and nanopillar heights required for high-performance devices\cite{Park2019}. Similarly, polarization converters written by bulk femtosecond lithography\cite{Baltrukonis2021,richter2012nanogratings,drevinskas2017high} cannot accurately shape each nanoelement, compromising their ability to encode complex spatially varying designs, and suffer from slow writing time, limiting their production volume for large area optics required in high-intensity experiments. While these approaches are monolithic, limitations in nanofabrication restrict their scalability or ability to fully control the optical response.

Here, we present a full-silica meta-optics technology based on high-aspect-ratio nanoelements that overcomes these restrictions. By coupling the intrinsic robustness and broadband transparency of silica with anisotropic etching, we unlock a full library of ultra-slender meta-atoms capable of delivering phase and polarization control while withstanding high-intensity laser operation. Our work sets the foundation for high-intensity structured light with unprecedented scalar and vectorial\cite{Rosales2018a} beam shaping, opening new opportunities in ultrafast science and technology requiring both high laser energies and finely sculpted optical modes.

\bigskip
\textbf{\large{Results}}

\begin{figure}[h]
\centering\includegraphics[width=0.8\textwidth]{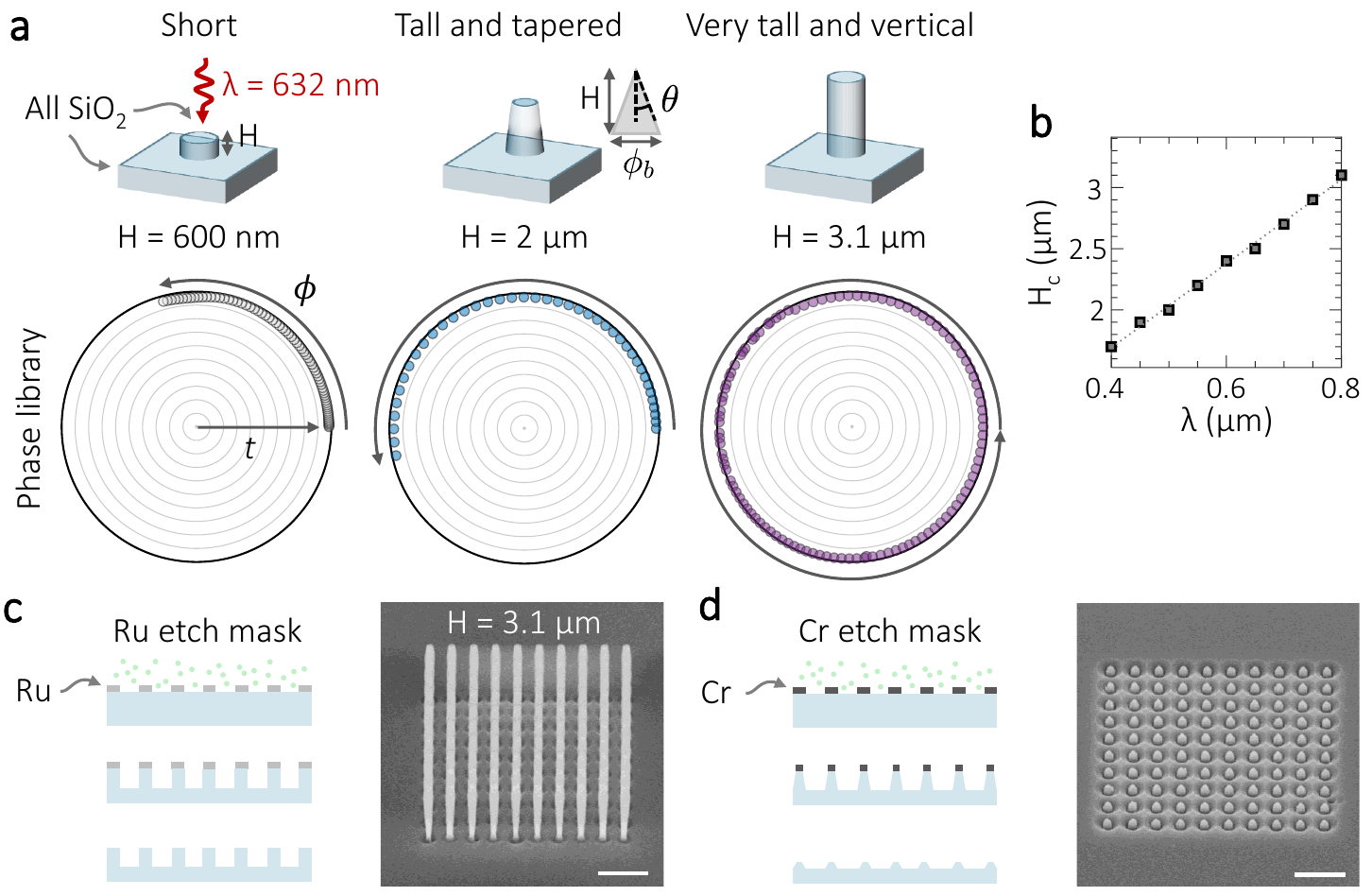}
\caption{\textbf{The importance of high-aspect-ratio nanopillars}---\textbf{a}, Phase library simulations for silica nanopillars of different height $H$ and sidewall verticality, on silica substrate.  The tapered case also shows a conical nanopillar with base diameter $\phi_b$ and taper angle ($\theta$). Depending on the diameter of each nanoelement, a different response is obtained and displayed on a circle of complex transmittance ($t$, amplitude transmission); $\phi$, transmitted phase---the absolute phase is shifted to start in each case from zero azimuthal angle). Illumination wavelength is 632 nm and unit cell size is 500 nm. A full phase library is obtained only for very tall and vertical nanopillars. \textbf{b}, Critical height to achieve a full phase library at different wavelengths. \textbf{c}, Schematic of the reactive ion etching and mask removal steps for silica and the ruthenium mask (left), and $45^\circ$ scanning electron microscope (SEM) image of the resulting nanopillars with $\sim14$ aspect ratio. \textbf{d}, Same as \textbf{c} but for a chromium mask resulting in shallow domes due to tapering and mask erosion. Scale bar in the SEM images is 2 $\mu$m.}
\label{fig:fig1_vertical_library}
\end{figure}

\textit{Design and nanofabrication}---The key challenges in using silica for meta-optics stem from its low refractive index ($n_{SiO_2}\approx1.45$ in the visible), which is nearly half that of high-index dielectrics commonly employed in metasurface designs (e.g. $n_{TiO_2}\approx2.6$). Because the index contrast between a silica nanopillar and air is small, the optical confinement is weak: resonant fields spill out more strongly, enlarging the mode volume and reducing the phase shift imparted by each nanoelement. By contrast, high-index nanopillars can support tightly confined resonant modes over subwavelength dimensions. Consequently, short ($\sim\lambda$ tall, where $\lambda$ is the illumination wavelength) silica nanopillars provide only a limited phase-shift range by varying their cross-section, resulting in narrower phase libraries (Fig. \ref{fig:fig1_vertical_library}a, left). 

Increasing nanopillar height to many wavelengths transforms each pillar into a vertical waveguide. Even with a low refractive index, the longer propagation distance inside the structure increases the total phase accumulation. Such tall designs, however, are challenging to fabricate as they often exhibit tapering\cite{Park2019}, which again restricts the overall phase coverage (Fig. \ref{fig:fig1_vertical_library}a, center). Moreover, the taper imposes a hard limit on the minimum base diameter $\diameter_b = 2H \mathrm{tan}(\theta)$, where $H$ is the height and $\theta$ is the taper angle ($\approx2.85^\circ$ in SiO$_2$ \cite{Park2019}), leading to conical pillars.

Ideally, one would have very tall, vertical nanopillars. Such a geometry allows one to rely on a standard cylindrical design---without resorting to corrective factors---while spanning a full phase library (Fig. \ref{fig:fig1_vertical_library}a, right). The wavelength dependence of the critical height $H_c$ to achieve a full phase library is nearly linear (Fig. \ref{fig:fig1_vertical_library}b). By replacing the conventional chromium etch mask with a ruthenium mask\cite{mitchell2021highly}, we demonstrate a reactive-ion etching process based on a fluorinated-gas/hydrogen mixture (CF$_4$/H$_2$) that enables high-aspect-ratio silica structures. We achieve cylindrical nanopillars 3.1 $\mu$m tall (corresponding to $\sim5\lambda$ at HeNe wavelength), with base diameters down to 220 nm and aspect ratios of $\sim14$ (Fig. \ref{fig:fig1_vertical_library}c), surpassing the most recent results in the literature based on thermal oxidation of silicon nanopillars \cite{okatani2023fabrication} ($8.7$ aspect ratio and 1.5 $\mu$m height). In comparison, using the same recipe with a chromium mask results in shallow dome-like shapes---an effect we attribute to mask erosion during reactive ion etching and collapse once $H>\diameter_b/(2 \mathrm{tan}\theta)$ (Fig. \ref{fig:fig1_vertical_library}d). 

\bigskip

\begin{figure}[h]
\centering\includegraphics[width=0.8\textwidth]{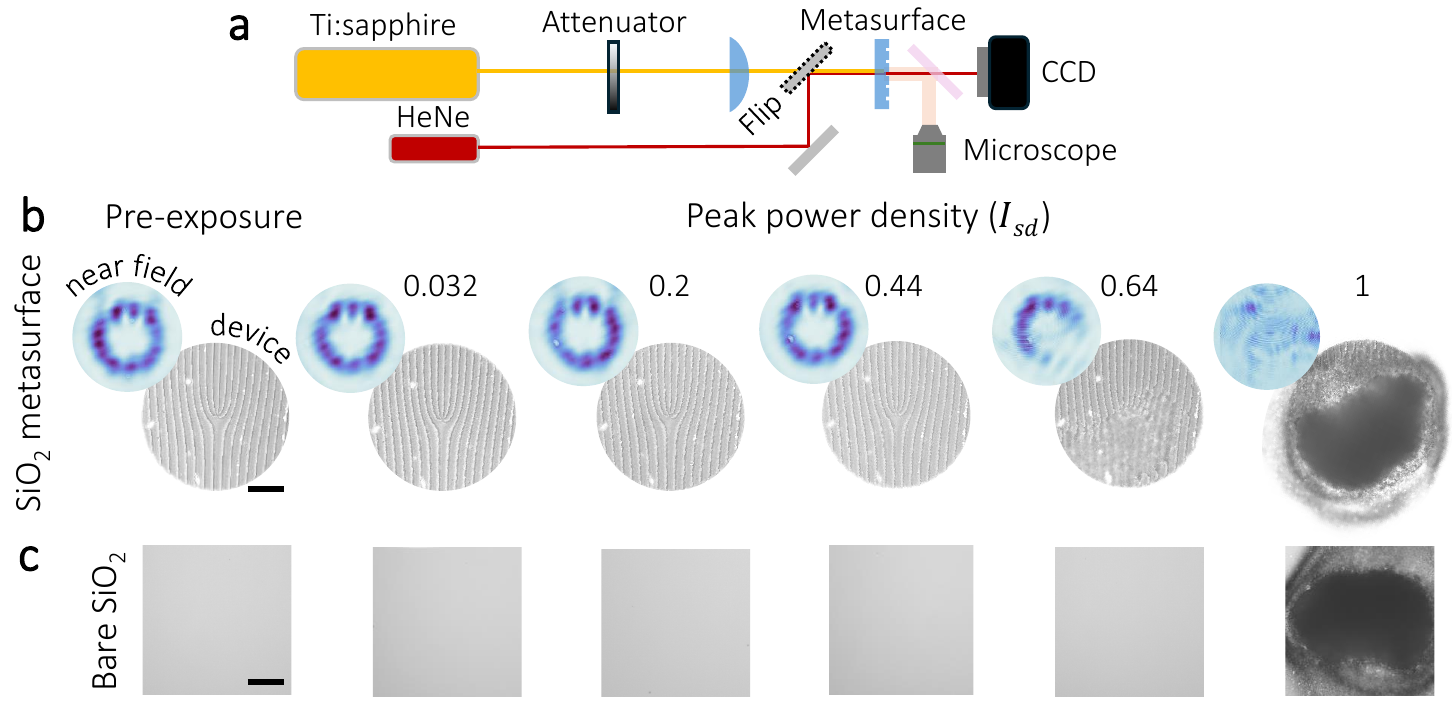}
\caption{\textbf{Robustness to high-intensity lasers}---\textbf{a}, Schematic of the laser-induced damage threshold setup. The high-intensity Ti:sapphire laser emits 40~fs pulses at 800 nm wavelength and 1 kHz repetition rate. The attenuator allows setting the peak power density focused on the metasurface by a lens. Both the near field beam structured using a HeNe probe laser and the surface image of the device are monitored using a camera (CCD) and a microscope, respectively. \textbf{b}, Near-field beam and surface image of the metasurface with 2 $\mu$m tall nanopillars before exposure and after exposure at various peak power density levels. The latter are normalized to the silica substrate damage threshold ($I_{sd}$). \textbf{c}, Surface image of the bare substrate before and after exposure. Scale bar is 25 $\mu$m in all devices.}
\label{fig2_LIDT_2um}
\end{figure}

\textit{Laser-induced damage threshold}---To assess the robustness of silica meta-optics at high intensities, we performed in-house laser-induced damage threshold (LIDT) measurements using a femtosecond laser. At each exposure level, we inspect the same metasurface device via optical microscopy and monitor its operational performance using a donut-shaped probe beam generated by the device (Fig. \ref{fig2_LIDT_2um}a). This metasurface is a vortex-generating optic, whose design and functionality are discussed in detail later. Here, we focus on its power handling capabilities under high-intensity illumination.

For reference, a bare silica region (Fig. \ref{fig2_LIDT_2um}c) was also exposed and used to normalize the peak power density to the substrate damage threshold ($I_{sd}$), which in our experiments corresponded to a peak power density of $5.8\cdot10^{13}$ W/cm$^2$ and a fluence of $2.3$ J/cm$^2$ at 40 fs long pulses, consistent with typical values of silica’s LIDT under these illumination conditions\cite{Mero2005}.
 
Notably, we find that the silica metasurface is resilient up to a sizeable fraction of the substrate’s damage threshold. For 2 $\mu$m-tall nanopillars, damage occurs at 0.64 $I_{sd}$  (Fig. \ref{fig2_LIDT_2um}b) whereas 3.1 $\mu$m-tall pillars sustain up to the range $0.2\text{-}0.39$ $I_{sd}$ (Fig. S1). These results indicate that increasing pillar height partially lowers the metasurface’s LIDT. Nevertheless, even for these record-high nanopillars and despite the limited range of our custom LIDT setup, the measured damage threshold of silica meta-optics is within a factor of $\sim6$ from bare silica, being one of the most robust materials for high-intensity lasers. Furthermore, identical tests performed on silicon-on-silica heterogeneous metasurfaces show damage already at the lowest power density level we studied (setting an upper bound of 0.03$I_{sd}$), underscoring the significantly improved resilience of all-silica meta-optics compared with conventional metasurfaces. 

\bigskip

\begin{figure}[h]
\centering\includegraphics[width=1\textwidth]{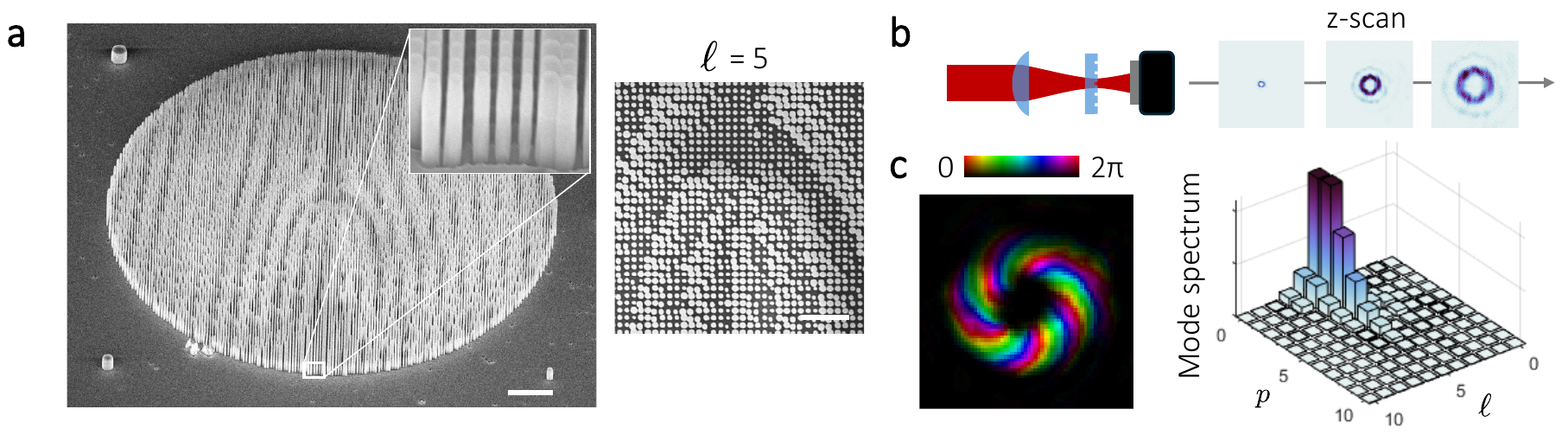}
\caption{\textbf{Full-silica vortex generator}---\textbf{a}, Tilted ($45^\circ$) scanning electron microscope (SEM) image (left) of the silica meta-optic for orbital angular momentum (OAM) generation with $\ell=5$ topological charge. Scale bar is 10 $\mu$m. Inset shows the $3.1 \mu$m tall nanopillars.  Also shown is a top view of the central region of the device (right) with scale of 4 $\mu$m. \textbf{b}, Schematic of the setup for the $z$-scan measurement enabling the wavefront reconstruction showing samples of intensity images of the donut beam measured at different propagation distances. \textbf{c}, Reconstructed complex amplitude of the structured beam (left), and corresponding modal decomposition spectrum on the radial ($p$) and azimuthal ($\ell$) modes of a Laguerre-Gaussian basis.}
\label{fig3_vortex}
\end{figure}

\textit{Vortex generation}---Having established the robustness of the silica meta-optics, we next demonstrate their phase-manipulation capabilities. We design a vortex generator implementing a pitchfork phase profile that combines the azimuthal component of a vortex with topological charge $\ell=5$ and a diffraction grating, realized with 3.1 $\mu$m tall nanopillars (Fig. \ref{fig3_vortex}a). The device operates at HeNe wavelength for convenience of our diagnostics. The resulting output beam displays the characteristic donut intensity profile, and we measure a modulation efficiency of 70\%---defined as the first-order diffracted power normalized by the total transmitted power---which is partly limited by deviations in nanopillar dimensions arising in the fabrication. To retrieve its phase structure, we perform a $z$-scan measurement of the intensity profile (Fig. \ref{fig3_vortex}b) and apply an iterative wavefront-reconstruction algorithm to the sequence of collected intensity images based on the Fresnel propagator \cite{pedrini2005wave,dorrah2024free} (Supplementary Material). The algorithm starts from a random initial phase profile, runs for 1000 iterations, and converges to an azimuthal phase gradient with five discrete sectors in excellent agreement with the metasurface design  (Fig. \ref{fig3_vortex}c, left).

From the reconstructed complex field, we perform a modal decomposition into Laguerre–Gaussian modes\cite{pinnell2020modal}, confirming that most of the optical power is concentrated in the $\ell=5$ modes as designed (Fig. \ref{fig3_vortex}c, right). The small spread in radial  ($p$)  modes mainly reflects the known limitation of purely phase-based elements, where the lack of amplitude modulation yields additional radial mode content\cite{Sephton2016}---an effect that phase–amplitude metasurfaces can eliminate\cite{Piccardo2020arbpol,Oliveira2023}. Additional contributions to the spread in $\ell$ and $p$ modes are the quadratic phase term introduced by the lens needed for the $z$-scan measurement, and deviations of nanopillar dimensions from design. This demonstration of a full-silica vortex metasurface opens a new route towards robust orbital angular momentum (OAM) generators in laser-matter interactions, taking advantage of the unparalleled control metasurfaces can offer at high intensities, such as propagation-dependent OAM\cite{dorrah2021structuring}, spin-controlled OAM\cite{devlin2017arbitrary}, and time-dependent OAM\cite{chen2022synthesizing}. 

\bigskip

\begin{figure}[h]
\centering\includegraphics[width=1\textwidth]{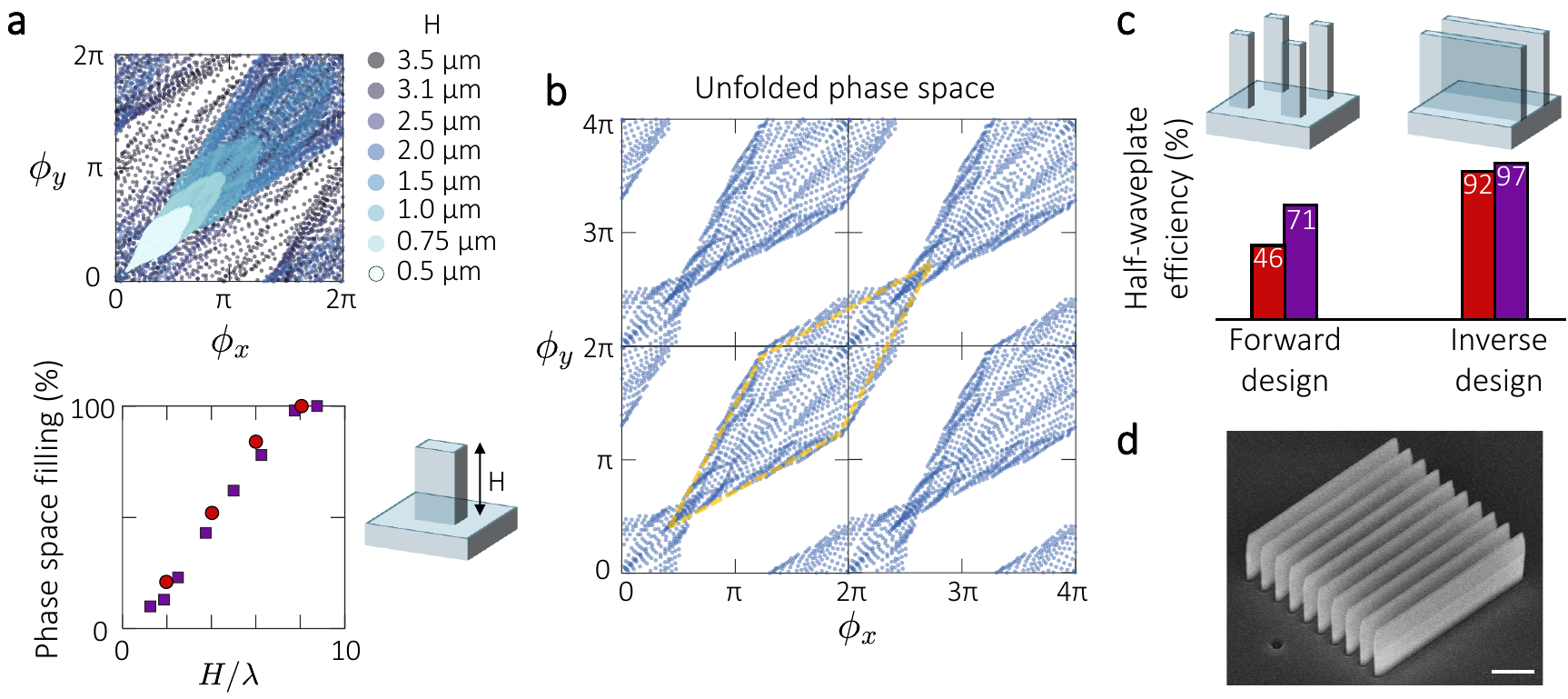}
\caption{\textbf{Meta-atom design for polarization control}---\textbf{a}, Phase library simulations of birefringent all-silica nanopillars with different height $H$ (top). Each simulated nanopillar has a different rectangular cross section and its phase shift imparted to $x$- and $y$-polarized light is displayed on the plot. Illumination wavelength is 400 nm and unit cell size is 390 nm. Also shown is the two-dimensional phase space filling of each library as a function of the nanopillar height normalized to illumination wavelength (bottom) for $\lambda=400$ nm (violet squares) and $\lambda=632$ nm (red circles). \textbf{b}, Phase library obtained for $H=2$ $\mu$m and $\lambda=400$ nm plotted in unfolded phase space. \textbf{c}, Simulated half-waveplate efficiency based on the optimal geometry of rectangular nanopillars (forward design) and that of nanogratings (resulting from inverse design topology optimization\cite{Keraly2013}). In both cases we calculate the efficiency for $\lambda = 400$ nm (violet bars) and 632 nm (red bars) illumination. \textbf{d}, Tilted ($45^\circ$) scanning electron microscope (SEM) image of the silica nanograting with $3.1$ $\mu$m height and $\sim$12 aspect ratio. Scale bar is $2$ $\mu$m.}
\label{fig4_polcontrol}
\end{figure}



\textit{Design of birefringent metasurfaces}---We now turn to the more demanding case of polarization control. In this scenario, covering the $0$ to $2\pi$ phase range for each of the two orthogonal polarization states renders the design space effectively two-dimensional. By simulating a range of nanopillar libraries of different heights, we find that beyond a critical $H_c$, the entire folded birefringent phase space is covered (Fig.~\ref{fig4_polcontrol}a, top and Fig. S2). This threshold occurs at $H_c/\lambda\sim8$ in both the near-ultraviolet and red portions of the spectrum (Fig.~\ref{fig4_polcontrol}a, bottom), given the scale invariance of Maxwell’s equations and that silica’s refractive index varies only slightly over this spectral range.

We observe that at low $H$ the phase response distributions of the nanopillars in the two-dimensional space resemble a set of rhombi with different diagonal lengths and nearly constant vertex angles. To gain insight into the mechanism of birefringent phase space filling, we examine the distribution for a specific height ($H=2 \mu$m) in the unfolded phase space (Fig. \ref{fig4_polcontrol}b). In this representation, the periodic boundary conditions of the folded phase space are removed, revealing that the rhombus begins to extend beyond the original periodic cell. At this height the phase space is not yet uniformly filled, however, the populated regions in the corners of the folded phase space indicate the onset of connectivity across boundaries.  These regions correspond to new phase retardance values emerging at larger $H$. As $H$ increases further, these corner regions grow, progressively covering more of the folded phase space, until complete birefringent phase space filling is achieved.

Near half-wave retardance ($\Delta\phi = |\phi_y -\phi_x|=\pi$) lies the most challenging region to fill. Given the fabrication complexities of maintaining vertical sidewalls at pillar heights on the order of multiple wavelengths ($H_c\sim8\lambda$ for large retardance), the nanopillar approach can prove inadequate, particularly in the visible. While we can fabricate 3.1 $\mu$m tall pillars---sufficient for near-ultraviolet coverage---this height remains suboptimal for the longer wavelengths in the visible and near-infrared range. To address this limitation, we implement a topology optimization algorithm \cite{sell2017large,shi2020continuous,fan2020freeform} (Methods) that seeks to maximize polarization-conversion efficiency in a half-wave plate. As shown in Fig.\ref{fig4_polcontrol}c, the efficiency of these inverse-designed structures far surpasses that of rectangular pillars for both near-ultraviolet and red wavelengths, all with a $3.1 \mu$m-tall design. Notably, the algorithm effectively rediscovers subwavelength nanogratings---employed long time ago for generating vector beams\cite{bomzon2002radially}---confirming that these are optimal half-waveplate structures under tight height constraints. Motivated by this finding, we fabricated silica subwavelength nanogratings (Fig.\ref{fig4_polcontrol}d), similar to earlier ultraviolet-range structures\cite{bonod2021full} but with over $3\times$ larger height and aspect ratio, further expanding the design range for high-intensity polarization optics.

\bigskip

\begin{figure}[h]
\centering\includegraphics[width=1\textwidth]{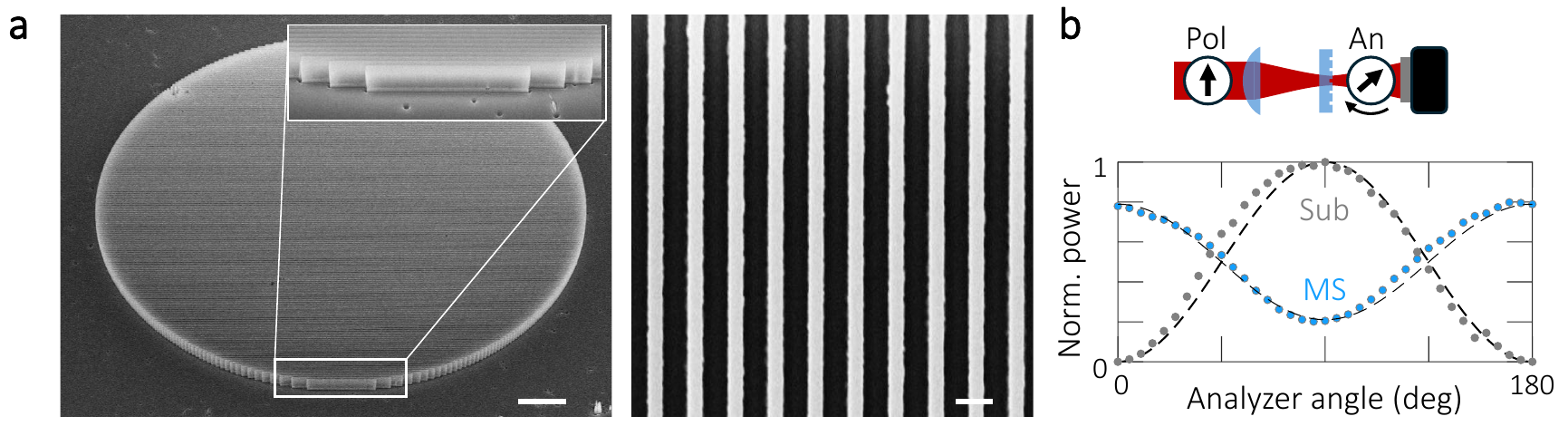}
\caption{\textbf{Full-silica waveplate with large retardance}---\textbf{a}, Tilted ($45^\circ$) scanning electron microscope (SEM) image of the silica waveplate (left) with 10 $\mu$m scale bar. Inset shows the 3.1 $\mu$m tall vertical nanostripes. Also shown is the top view SEM image of the stripes with 500 nm scale bar. \textbf{b}, Schematic of the polarimetry setup (top). A fixed polarizer (Pol) is aligned at $45^\circ$ with respect to the waveplate's fast axis. An analyzer (An) is rotated and the transmitted power through the system is integrated by a power meter. The curves measured by transmitting laser light through the bare substrate (Sub) and through the metasurface (MS) are also shown (bottom). Illumination wavelength is 632 nm.}
\label{fig5_waveplate}
\end{figure}

\textit{Phase retardance of polarization meta-optics}---To assess the performance of our high-aspect-ratio silica waveplates we fabricate a set of subwavelength nanograting devices with 3.1 $\mu$m height (Fig. \ref{fig5_waveplate}a). We measure them implementing a rotating-polarizer polarimetry setup\cite{williams1997rotating,zhang2013methods} (Methods). At the HeNe wavelength, the largest phase retardance we measure is $125^\circ$. The corresponding transmitted power curve as a function of analyzer angle appears out-of-phase with respect to bare silica, as expected (Fig. \ref{fig5_waveplate}b). This device can convert horizontal to vertical polarization with $80\%$ efficiency. Several additional waveplates with phase retardance down to $20^\circ$ at HeNe wavelength were also successfully produced. Further refinements in fabrication fidelity and increases in nanograting height promise to boost half-waveplate performance at visible wavelengths, while operation at shorter, ultraviolet wavelengths should readily achieve half-wave retardance, far surpassing previous demonstrations of quarter-waveplate metasurfaces in that spectral range\cite{Ray2023pol,bonod2021full}.

\bigskip
\textbf{\large{Discussion}}

Our demonstration of high-aspect-ratio silica meta-optics offers a key step forward in delivering complex beam shaping at high intensities. By combining silica's high LIDT and anisotropic etching processes, we achieve meta-optics with damage thresholds comparable to that of the substrate. The addition of anti-reflection coatings\cite{Ray2020} and further improvements in the nanofabrication fidelity is expected to bring their efficiency above 90$\%$. Scaling the size to centimeter-diameter metasurfaces will allow shaping femtosecond pulses at the joule level. Such devices are expected to handle peak power densities of $10^{13}$ W/cm$^2$ and with subsequent focusing of pulses to 10 $\mu$m spots---chosen to preserve beam structure rather than reach the diffraction limit---will reach relativistic intensities on the order of 10$^{19}$ W/cm$^2$. Even reducing the device size to the millimeter scale would still allow intensities around 10$^{17}$ W/cm$^2$ at focus---being in the plasma ionization regime.

These new capabilities unlock a host of structured high-intensity laser–matter interaction scenarios that stand to benefit from the powerful design freedoms inherent to metasurfaces. For example, it will be possible to generate vortex beams with large OAM in extreme conditions enabling relativistic OAM-driven particle acceleration schemes\cite{Vieira2018,willim2023proton}. Moreover, because metasurfaces permit spin-dependent or intensity-tunable transformations\cite{Dorrah2022}, the same device architecture could be adapted to vary beam profiles on demand, significantly increasing the agility of ultrafast experiments. Beyond OAM, silica meta-optics may find use in imparting arbitrary phase corrections to counteract aberrations in high-power laser facilities, reducing bulk-optic requirements. Their compact form factor and planar geometry also render them attractive for emerging nonlinear-optical experiments, such as high-harmonic generation from structured pump beams, where precise spatial and temporal control is crucial.

Our polarization control results hold the promise of extending the paradigm to vectorial beam shaping, using conventional nanopillars in the near-ultraviolet and short visible wavelengths, and spatially bending nanograting elements \cite{bomzon2002radially} at longer wavelengths. Finally, just as multimode photonics is revolutionizing the design of spatiotemporal pulses at moderate power\cite{Froula2018,Piccardo2023broadbandcontrol,Yessenov2022}, extending these concepts to the high-intensity domain could unlock a new era in ultrafast science, where simultaneously sculpting spatial, spectral, and temporal degrees of freedom may lead to unprecedented control over light–matter interactions, from tabletop ultrafast intense sources to large-scale laser facilities.









\bigskip
\textbf{\large{Funding}}---We acknowledge funding of the Research Unit INESC MN from the Fundação para a Ciência e a Tecnologia (FCT) through the BASE (UIDB/05367/2020) and PROGRAMATICO (UIDP/05367/2020) programs, CTI (Centro de Tecnologia e Inovação)---Missão Interface/PRR (RE-C05-i02). M.P. received funding from the European Research Council (ERC StG) under the European Union’s Horizon Europe research and innovation program (Grant agreement No. 101161858). M.F., P.S.M.C. and P.E. acknowledge funding from  European Innovation Council Pathfinder Project 101047223 NanoXCAN.

\bigskip
\textbf{\large{Acknowledgments}}---The authors gratefully acknowledge INESC MN for providing the cleanroom facilities. M.P. thanks J.-S. Park and A. Palmieri (Harvard) for useful discussions, and D. Kazakov for a careful reading of the manuscript.

\bigskip
\textbf{\large{Disclosures}}---The authors declare no conflicts of interest.

\bigskip
\textbf{\large{Data Availability}}---Data underlying the results presented in this paper may be obtained from the authors upon reasonable request.



\bibliography{highintensitymeta}






\end{document}